# *Implementing New-age Authentication Techniques using OpenID for Security Automation*

*Dharmendra Choukse*
Institute of Engg. & Science,
IPS Academy,
Indore, India

*Umesh Kumar Singh*
Institute of Comp. Science,
Vikram University,
Ujjain, India

*Deepak Sukheja*
Priyatam Institute of Tech.
and Managemnet,
Indore, India

*Rekha Shahapurkar*
Lokmanya Tilak College,
Vikram University,
Ujjain,India

**Abstract**

*Security of any software can be enhanced manifolds if multiple factors for authorization and authentication are used .The main aim of this work was to design and implement an Academy Automation Software for IPS Academy which uses OpenID and Windows CardSpace as Authentication Techniques in addition to Role Based Authentication(RBA) System to ensure that only authentic users can access the predefined roles as per their Authorization level. The Automation covers different computing hardware and software that can be used to digitally create, manipulate, collect, store, and relay Academy information needed for accomplishing basic Operation like admissions and registration , student and faculty interaction, online library, medical and business development. Raw data storage, electronic transfer, and the management of electronic business information comprise the basic activities of the Academy automation system. Further Transport Layer Security (TLS) protocol has been implemented to provide security and data integrity for communications over networks. TLS encrypts the segments of network connections at the Transport*

## 1. INTRODUCTION

The World Wide Web (WWW) is a critical enabling technology for electronic commerce on the Internet. Its underlying protocol, HTTP (Hypertext Transfer Protocol [Fielding et al. 1999]), has been widely used to synthesize diverse technologies and components, to great effect in Web environments. Increased integration of Web, operating system, and database system technologies will lead to continued reliance on Web technology for enterprise computing. However, current approaches to access control on Web servers are mostly based on individual user identity; hence they do not scale to enterprise-wide systems. If the roles of individual users are provided securely, Web servers can trust and use the roles for role-based access control (RBAC [Sandhu et al. 1996; Sandhu 1998]). So a successful marriage of the Web and RBAC has the potential for making a considerable impact on deployment of effective enterprise-wide security in large-scale systems. In this article present a comprehensive approach to RBAC on the Web to identify the user-pull and server-pull architectures and analyze their utility. To support these architectures on the Web, for relatively mature technologies and extend them for secure RBAC on the Web. In order to do so, to make use of standard technologies in use on the Web: cookies [Kristol and Montulli 1999; Moore and Freed 1999], X.509 [ITU-T Recommendation X.509 1993; 1997; Housley et al. 1998], SSL (Secure Socket Layer [Wagner and Schneier 1996; Dierks and Allen 1999]), and LDAP (Lightweight Directory Access Protocol [Howes et al. 1999]).

The Lightweight Directory Access Protocol (LDAP) directory service already available for the purpose of webmail authentication of IPS Academy, Indore users has been used to do the basic Authentication. The client can request the application server for any web application which will ask for the user credentials which will be verified in the LDAP server through an ASP.Net [19] Module. On successful verification, the authorization module will contact the user role database and fetch the roles for that user. In case of return of multiple roles, user will be given the authorization of all the roles. The access to the application will be on the basis of privilege of the role of that particular user. The role database is implementing in Microsoft SQL server [18] database.

On successful authentication, the Authentication and authorization module which has been developed for this purpose is called and the role for the user is retrieved. Privileges are granted to roles which in turn roles are granted to users. The overall database server and application server is considered for possible attacks. The proposed scheme is given in figure 3. The database server and the authentication server are in a private network and separated from the user network by a firewall. These servers can be accessed only through application server, i.e. through the authentication and authorization module. Application server has an interface in the private network but can avail only the specific service which has been explicitly allowed in the firewall. Application server has another interface which is part of user network with a firewall to restrict the clients only to the desired service.

## 2. OBSERVATION AND PROBLEM DESCRIPTION

The whole Collage Academy automation consists of many sections viz. Student Affairs, Academic Section, Research and development, Training and Placement, Finance and Accounts be given access to different aspects of the systems based on their clearance level. For e.g. the Assistant Registrar of Student Affairs should have full access to all the options of Student Affairs database but not that of the Academic Section database. However, provisions have to be made so that he/she is able to perform some student affairs related queries to the student affairs database. Similarly, a student must have read-only access to his/her information in the official records and modifying capabilities some of his/her details in the training and placement section database. This calls for a role-based approach to access the databases. Each person has a certain role attached to it. This role corresponds to the areas of the work his login account can access. If a violation occurs, the user is immediately logged out.

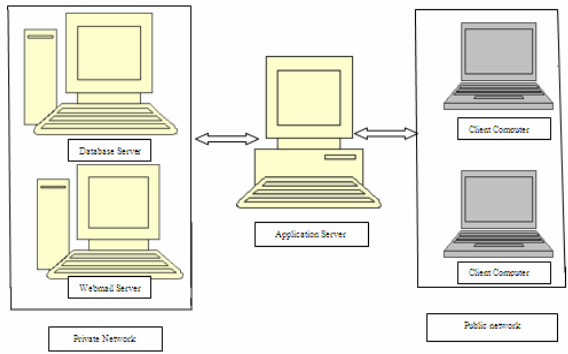

Figure 1: System and Server Security

In this work the design and implementation of the Role Based Authentication Schemes for Security Automation is described, developed at the IPS Academy, Indore as an ASP.NET 2005 web application in C# server side code, HTML, and JavaScript for use on the Internet. The purpose work to deploy a cost-effective, web-based system that significantly extends the capabilities, flexibility, benefits, and confidentiality of paper-based rating methods while incorporating the ease of use of existing online surveys and polling programs.

2.1 *Problem Issues and Challenges*

There are Following problems:-
1. The information line must be completely secured.
2. Proper Encryption must be used for storing the Password for the User.
3. The authorization token which is stored on the client side has to be encrypted so that the client cannot modify his authorization clearance level.
4. Each userid-role mapping should have an expiry date beyond which it will be invalid.
5. Role Scoping: Local and Global Roles
6. In each role, we have to have an owner. Normally the role will map to the user id of the owner. The owner can change the mapping and can specify the time period of this change. The newly mapped user is not the owner and so cannot change the ownership, but maybe allowed to map again. For example, HODCSE is the role and the owner's user id is"Ram". Normally, HODCSE maps to Ram. When Prof.

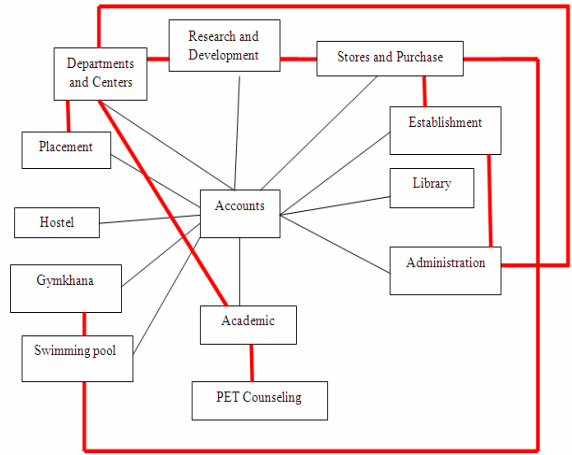

Figure 2: Basic Architecture of Academy

Ram goes on leave, he fills up some form electronically and this triggers (among other things) a role change of HODCSE to the user he designates, say Prof.Shayam. Now"Ram" is going on leave till 4/7/2009, so the changed mapping is till 4/7/2009 (to"pshayam"; specified by"Ram" in the form he filled up). Now due to an emergency, "pshayam" had to leave station on 4/7/2009, making Prof manoj the Head. Since" pshayam" is not the owner, he cannot change the validity date beyond 4/7/2009 and"Ashish" takes over the HODCSE role till 4/7/2009. On 5/7/2009 (or the next query of the role), the role remaps to"Ram". Other cases (like"Ram" having to overstay beyond 4/7) can be handled by the administrator

3. METHODOLOGY

1. We have 2 sets of Roles:
- **Global Roles**: These refer to the roles which are common to all the applications viz. root,
   Director: Their Role IDs are of single Digit: 0, 1, and 2 etc.
- **Local Roles**: These are roles which are specific to a module. For E.g. for Student Affairs,the roles of Assistant Registrar, Academy in charge. Their IDs are of the Form: 10, 11, 12 ... 110 etc. where first digit identifies the application to which all of them are common.
2. There is a Global role to role_id mapping table.
3. Also there is a local mapping table for each section. Insertion/modification or deletion of any entry in the local table generates a Microsoft SQL trigger for its 'encoded' entry addition in the global table.

A web interface which is accessed by any member and is used to assign his role to any other member for a specified period. The role validity period of the other person cannot exceed the validity period of the assigner. So, whenever a role has to be transferred, an entry is made in the user role relation table corresponding to the user ID of the assigned person and it is made sure that the validity period of the assigned is less than the validity period of assigner from the same user role relation table.

TABLE 1: Various roles and their IDs

| Role | Role ID |
|---|---|
| Administrator | 0 |
| Student | 1 |
| Faculty | 2 |
| Assistant Registrar (Student Affairs) | 10 |
| Assistant Registrar (Academic) | 20 |
| Assistant Registrar (RND) | 30 |
| Assistant Registrar (TNP) | 40 |
| Assistant Registrar (Finance) | 50 |
| Registrar | 3 |
| Director | 4 |
| Head of Departments | 5 |

TABLE 2: User name id relation

| User_name | User_id |
|---|---|
| root | 1 |
| dharmendra | 2 |
| try | 3 |

TABLE 3: User role relation

| s_no | user_id | role_id | valid_from | valid_upto |
|---|---|---|---|---|
| 1 | 1 | 12 | 2008-01-01 | 2009-01-02 |
| 2 | 1 | 13 | 2008-01-01 | 2008-05-06 |
| 3 | 2 | 12 | 2007-01-01 | 2008-01-01 |

*3.1 Database Table Structure*

We will have a common login page for all the sections of the Academy Automation. The looks up table of the corresponding IDs are shown in table 1, 2 & 3.

## 4. ADDING NEW AGE AUTHENTICATION TECHNIQUES AND MECHANISM

*4.1 OpenID*

It is an authentication system that is based on the premise that anyone can have a URL (or alternatively an Extensible Resource Identifier (XRI) [7] which is allowed in version 2.0) and an OpenID Identity Provider (OP) which is willing to speak on behalf of this URL or XRI. During its short lifetime, OpenID has evolved through three versions, namely Open ID v1.0, v1.1 [5] and v2.0 [4]. Whilst the first two versions were only concerned with authentication, v2.0 has added the capability for the exchange of identity attributes as well [6]. The first version of OpenID (v.1.0) had several security weaknesses some of which were quickly patched in v1.1 (e.g. messages could be replayed), while others were fixed in v2.0. However, as described below, v2.0 still suffers from several security weaknesses, which may or may not pose a significant risk, depending upon the application that one wishes to secure with OpenID.

OpenID works as follows in Figure 3. When a user contacts a Service Provider (SP) that supports OpenID, he presents his URL (or XRI), and the SP contacts the URL to see who is the OP that speaks for it. This is the process of Identity Provider Discovery, and it bypasses the need for the Where Are You From service of Shibboleth and the Identity Selector in CardSpace. When XRIs are used, these similarly are resolved by the SP to find the OP that can authenticate the user. Once the identity provider has been discovered, the SP must establish a shared secret with it so that future messages can be authenticated, using the well known process of message authentication codes (MAC). The OpenID specifications use Diffie-Hellman to establish the shared secret between the OP and SP; unfortunately, Diffie-Hellman is vulnerable to man in the middle attacks. Once the OP and SP have established a shared secret, the SP redirects the user to the OP, to be authenticated by any mechanism deemed appropriate by the OP. During the authentication process the OP is supposed to check that the user wants to be authenticated to this SP, by displaying the "realm" of the SP to the user. (The realm is a pattern that represents part of the name space of the SP e.g. *.kent.ac.uk). But this realm information can easily be spoofed by an evil SP, which will lull the user into a false sense of security that they are authenticating to a known SP when in fact they will be redirected to an evil one after the authentication has completed. After successful user authentication, the OP redirects the user back to the SP along with an authentication token saying that the user has been authenticated and has control over the OpenID they specified. The SP then grants the user access to its services, as it deems appropriate. One might regard OpenID as a direct competitor of Shibboleth for user authentication. On the face of it OpenID sounds like an attractive way of assigning people globally unique IDs based on URLs, and an authentication service that will validate this binding. Unfortunately, when one inspects the OpenID system closer, one finds it has a significant number of weaknesses that Shibboleth does not have (and one major one that is shared with Shibboleth – namely Phishing).

*4.2 Advantages of OpenID*

*4.2.1 Simplicity*

The latest OpenID specification [4] is considerably thinner and much easier to understand than the latest SAML specification [11]. This is due to their scope. OpenID is concrete and specifies how data is carried over HTTP. SAML is an abstract framework and requires profiles and bindings to specify what content is carried over which Internet protocols. Thus the OpenID specification is complete and self contained (if one excludes its optional extensions), whereas SAML is a framework that references more than a dozen other specifications, which need to be understood in order to fully appreciate and implement SAML. To secure messages, SAML relies on XML, and consequently upon XML signatures and XML encryption, whereas OpenID simply relies on SSL/TLS. Whilst both specify a protocol for carrying authentication assertions between users and various identity and service providers, OpenID's protocol is concrete whilst SAML's protocol is abstract and also includes attribute assertions and authorisation assertions. Whilst OpenID specifies the format of user identifiers and how the user's OpenID provider is discovered, SAML's framework does not mandate any particular user identifier, nor how the user's identity provider is discovered. Thus SAML is far more flexible and extensible than OpenID, but as a result, is more complex. However, it would be possible to define a profile of SAML and a protocol binding that would mirror that of OpenID if this was required.

*4.2.2 Implementation Effort*

An OpenID infrastructure should be significantly easier to implement and deploy than a SAML-based one. Because the OpenID specification is simpler, with far fewer options, most OpenID implementations should interwork "out of the box". SAML implementations on the other hand will most likely only implement a specific subset of profiles and protocol bindings, which will necessarily mean that not all implementations will interwork "out of the box", and may require significant configuration in order to do so.

*4.2.3 No Discovery Service*

Because OP discovery is built into the OpenID specification, via the user's URL or XRI, then there is no need for a separate Discovery mechanism. Shibboleth on the other hand does require a means of IdP discovery which can be provided in a number of ways (though none of these is satisfactory as a long-term solution). As a default, the federation provides a Where Are You From (WAYF) service, which presents users with a list of IdPs from which to choose; this becomes increasingly less viable as the number of IdPs grows. Institutions which have implemented library portals have been able to improve on this by providing their users with direct links to the institution's subscribed services (so called WAYFless URLs); in the Schools sector this method of service presentation is seen as essential. Another solution is provided by some large publishers and JISC data centres, which present the user with a list of institution names drawn from the service's list of subscribers (though for SPs with a large client list this provides little relief from the current WAYF). Multiple WAYFs are not a realistic solution for supporting inter-federation connections though no alternative solution has been agreed. Note that the OpenID method of discovery is not new to academia. The EduRoam service [12] works in a similar way to OpenID by requiring the user to enter his identity in the form username@realm, where realm is the DNS name of the user's university, and acts as the discovery service for EduRoam. CardSpace on the other hand, relies on the user's PC to act as the discovery service. The Identity Selector program allows the user to pick the IdP he wants to use from the cards in his selector.

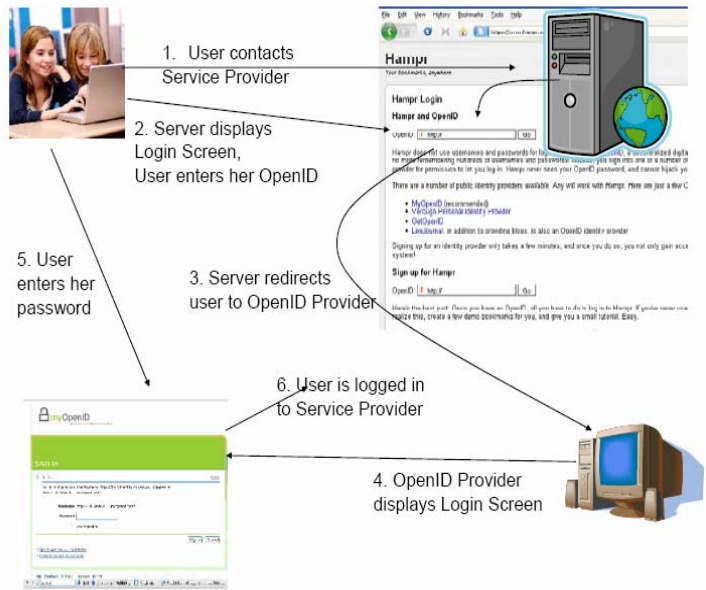

Figure 3:- The OpenID Authentication Process

*4.3 OpenID Implementation:*

To implement OpenID in IPS Automation System we have used JanRain OpenSource Library .The auto -mation application will go through authentication process as follows:

- End User enters his identity URL on the Login page.
- Create OpenID authentication request from User's URL.
- Initialize a storage place where the application can store information, since it is working in smart mode(to keep the session information).
- The request is sent to OpenID server for authentication. The request also includes the redirect path to the IPS Website.
- Response from the server is received and processed. In case of authentication failure , browser is redirected to an error page displaying a message that authentication failed. Otherwise the user will be logged in.

In the JanRain Library the "Services" folder contains the Yadis protocol files and "Auth" folder contains main OpenID library files( should be included in PHP search path). The record for ipsacademy.com is created in DNS and the web server(Apache) is configured to make JanRain library files become accessible when a page request is received . This is done as follows:

```
<VirtualHost *:80>
ServerAdmin webmaster@ipsacademy.org
DocumentRoot /backup/ipsacademy
ServerName ipsacademy.org
   ErrorLog logs/ipsacademy.org-error_log
```

```
    CustomLog logs/ipsacademy.org-access_log common
     </VirtualHost>
```
HTML source code segment for the OpenID Login form is as follows:
```
 <form method="get" action="try_auth.php">
 Identity URL:
 <input type="hidden" name="action" value="verify" />
 <input type="text" name="openid_url" value="" />
 <input type="submit" value="Verify" />
 </form>
```
The flow of authentication can be explained as: "index.php" page gets the OpenID URL from an End User and sends it to "try_auth.php" file. This file then constructs an authentication request and sends it to the OpenID server. The OpenID server processes this request and then sends the result back to "finish_auth.php" file using web browser redirection method.

*The index.php File*

This file is responsible asks user to enter his OpenID URL. The source code for this form is as shown below:
```
    <form method="get" action="try_auth.php">
    Identity URL:
    <input type="hidden" name="action" value="verify" />
    <input type="text" name="openid_url" value="" />
    <input type="submit" value="Verify" />
    </form>.
```

*The try_auth.php File*

This file is responsible for creating OpenID request and sending it to the OpenID server.

```
<?php
require_once "common.php";
session_start();
// Render a default page if we got a submission without an openid
// value.
if (empty($_GET['openid_url'])) {
$error = "Expected an OpenID URL.";
include 'index.php';
exit(0);
}
$scheme = 'http';
if (isset($_SERVER['HTTPS'])and $_SERVER['HTTPS'] == 'on') {
$scheme. = 's';
}
$openid = $_GET['openid_url'];
$process_url=
sprintf("$scheme://%s:%s%s/finish_auth.php",
$_SERVER['SERVER_NAME'],
$_SERVER['SERVER_PORT'],
dirname($_SERVER['PHP_SELF']));
$trust_root = sprintf("$scheme://%s:%s%s",
$_SERVER['SERVER_NAME'],
$_SERVER['SERVER_PORT'],
dirname($_SERVER['PHP_SELF']));
// Begin the OpenID authentication process.
$auth_request = $consumer->begin($openid);
// Handle failure status return values.
```

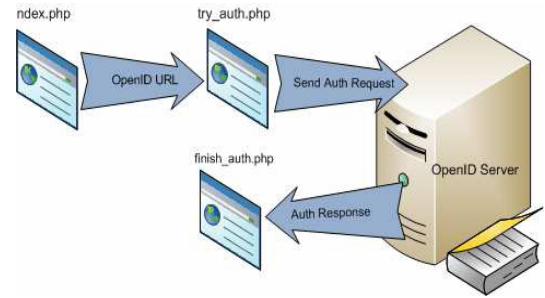

Figure 4: Use of ipsacademy source files during authentication request processing.

```
if (!$auth_request) {
$error = "Authentication error.";
include 'index.php';
exit(0);
}
$auth_request->addExtensionArg('sreg', 'optional', 'email');
// Redirect the user to the OpenID server for authentication. Store
// the token for this authentication so we can verify the response.
$redirect_url = $auth_request->redirectURL($trust_root, $process_url);
header("Location: ".$redirect_url);
?>
```

*The finish_auth.php File*

This file confirms successful verification or unsuccessful verification:

```
<?php
require_once "common.php";
session_start();
// complete the authentication process using the server's response.
$response = $consumer->complete ($_GET);
if ($response->status == Auth_OpenID_CANCEL) {
// This means the authentication was cancelled.
$msg = 'Verification cancelled.'
} else if ($response->status == Auth_OpenID_FAILURE)
{
$msg = "OpenID authentication failed: " . $response->message;
} else if ($response->status == Auth_OpenID_SUCCESS) {
// this means the authentication succeeded.
$openid = $response->identity_url;
$esc_identity = htmlspecialchars($openid, ENT_QUOTES);
```

```
$success = sprintf('You have successfully verified '
'<a href="%s">%s</a> as your identity.'
$esc_identity, $esc_identity);
if ($response->endpoint->canonicalID) {
$success .= ' (XRI CanonicalID: '.$response->endpoint-
>canonicalID.') ';
}
$sreg = $response->extensionResponse('sreg');
if (@$sreg['email']) {
$success .= " You also returned '".$sreg['email']."' as your
email.";
}
}
include 'index.php';
?>
```

After successfully authenticating with the OpenID provider (myopenid.com), the OpenID provider redirects the client to the page originally requested (http://<ipsacademy.org>:8081) that shows the remote user, groups, and roles of which the user is a member. The remote user is the openid identity that was used to log in. The groups contain a single group, the VALID_OPENID_USER group, and the role is populated with OPENID_ROLE. If this is not the case, the user would not be authenticated because only users in the role OPENID_ROLE are permitted to access this resource.

## 5. RECCOMMENDATION

- The IPS Automation System should keep track of both OpenID and CardSpace identity management systems as they evolve. There is clearly a great demand for a ubiquitous secure identity management system.
- Now that a publicly available OpenID gateway has been built, publicise its availability to the community and monitor its applications and usage. If usage becomes substantial, consider productising the service.

## 6. CONCLUSION

The research problem and goal of the Academy Automation is to design a highly secure and efficient framework based on Service Oriented Architecture. Keeping in mind the policies of minimum data redundancy and efficient security, the work revolved around designing a plug in for secure role based authentication. Presently the authentication is based on the traditional userid-password based approach, but as is suggested in this report, work can be done to incorporate new-age technologies such as OpenID. OpenID provides increased flexibility for application deployment by enabling applications to leverage and third-party authentication providers for handling authentication. Providers such as OpenID have become very common as more users want a single user profile across multiple sites for blogs, wikis, and other social networking activities. Additionally, many Web sites do not want to maintain, or require users to continually provide, the same profile-related information just to ensure that the user credentials are valid.

## REFERRENCE

AUTHORS PROFILE

**Biographical notes:**

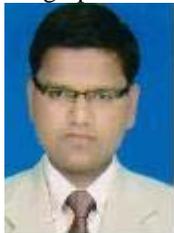

Dharmendra Choukse holds a M.Tech in Information Technology from Devi Ahilya University, Indore-INDIA. He is currently Pursuing Ph.D. in Computer Science From Institute of Computer Science, Vikram University, Ujjain-INDIA.and He is also currently Sr Software Engineer in Institute of Engineering & Sciences,IPS Academy, Indore-INDIA. A He served as Software Engineer in Choksi Laboratories ltd,Indore. His research interest includes network security, secure electronic commerce, client-server computing and IT based education.

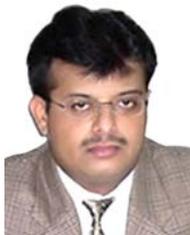

Dr. Umesh Kumar Singh obtained his Ph.D. in Computer Science from Devi Ahilya University, Indore-INDIA. He is currently Reader in Institute of Computer Science, Vikram University, Ujjain-INDIA. He served as professor in Computer Science and Principal in Mahakal Institute of Computer Sciences (MICS-MIT), Ujjain. He is formally Director I/c of Institute of Computer Science, Vikram University Ujjain. He has served as Engineer (E&T) in education and training division of CMC Ltd., New Delhi in initial years of his career. He has authored a book on " Internet and Web technology " and his various research papers are published in national and international journals of repute. Dr. Singh is reviewer of International Journal of Network Security (IJNS), ECKM Conferences and various Journals of Computer Science. His research interest includes network security, secure electronic commerce, client-server computing and IT based education.

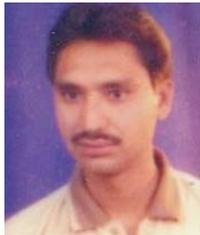

Deepak Sukheja holeds M.Sc.,M.Tech.NIT( Govt. Engg. College) Raipur-INDIA. He is currently Pursuing Ph.D. in Computer Science From Institute of Computer Science, Vikram University, Ujjain-INDIA and He is working as a Reader in Priyatam Institute of Technology and Managemnt Indore. He served as Sn. Software Engineer in Patni Compute System Mumbai, KPIT Pune and Tanmay Software Indore. His research interest includes network security, secure electronic commerce, client-server computing and Query Optimization.

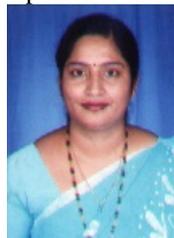

Rekha D Shahapurkar holds a MCA from Indira Gandhi National Open University, New Delhi, INDIA. She is currently Pursuing Ph.D. in Computer Science From Institute of Computer Science, Vikram University, Ujjain-INDIA. From 2001 she is working as Asst. Professor in Lokmanya Tilak College Ujjain-INDIA. Her research interest includes network security, client-server computing and IT based education.